\def\ln{\ell{n}}

\documentstyle[12pt,aps]{revtex}

\begin{document}
\draft
\title{On the estimation of the current value of the cosmological
constant}
\author{V.G.Gurzadyan$^{1,2}$ and She-Sheng Xue$^{1}$
}
\address{(1)
ICRA, INFN  and
Physics Department, University of Rome ``La Sapienza", 00185 Rome,
Italy;\\
(2) Yerevan Physics Institute and Garni Space Astronomy Institute,
Armenia;
}


\maketitle

\begin{abstract}

We advance the viewpoint that, only relevant modes of 
the vacuum fluctuations, namely, with wavelengths conditioned by the
size, 
homogeneity, geometry and topology of the Universe, do contribute into
the cosmological constant. 
A formula is derived which relates the cosmological 
constant with the size of the Universe and the three 
fundamental constants: the velocity of light, Planck and Newton
gravitational constants. 
Then the current value of the cosmological constant remarkably agrees 
with the value indicated by distant supernovae observations, 
i.e. is of the order of the
critical density. Thus the cosmological constant had to be smaller
than the matter density in the past and will be bigger in the future.

\end{abstract}



\narrowtext

\section{Introduction}

There are observational indications for the acceleration of
the expansion of the Universe\cite{sn} and that $\Omega_\Lambda$ 
attributed to the vacuum-energy density is of the order of the
critical density of the Universe. 
Though the accuracy of observational data is still a subject for
further
analysis, these observations 
have greatly revived the interest in the long-standing problem of the 
cosmological constant for not only its small value, but also its
closeness to the critical density of the Universe.

The cosmological term ($\Lambda$-term) $\Lambda g_{\mu\nu}$ was
introduced
by Einstein to incorporate the General Relativity with the Mach
principle.
To interpret apparently observed behavior of redshift distribution of
quasars 
within cosmological models of 
the $\Lambda$-term, Zeldovich\cite{zeld} revealed the origin of the
$\Lambda$-term
being attributed to the zero-point energy of the vacuum: the vacuum 
energy-momentum tensor $\tilde T_{ik}=\Lambda g_{ik}$ and the negative
vacuum pressure 
$\tilde p$ relating to the vacuum-energy density 
$\tilde T_{00}=\tilde \rho =-\tilde p$.

In the context of local quantum field theories in a flat space, the
vacuum 
fluctuations of various quantum fields result in 
a non-vanishing vacuum-energy density given by 
the spectrum $\epsilon_{B,F}(|{\bf k}|)$ of all quantum bosonic
and fermionic fields 
\begin{equation}
\tilde \rho={1\over2V}\sum_{\bf k} d^B_{\bf k}\epsilon_B(|{\bf k}|)
-{1\over V}\sum_{\bf k} d^F_{\bf k}\epsilon_F(|{\bf k}|),
\label{0}
\end{equation}
in a given volume of 3-dim space and 
summed over all possible ${\bf k}$ states with degeneracy
$d^B_{\bf k}$ and $d^F_{\bf k}$. It cannot be precluded the
contribution of the quantum gravity in eq.(\ref{0}).

In the four dimensional flat space-time $R^4$ and the
continuous spectrum $\epsilon_{B,F}(|{\bf k}|)$ of free quantum fields
with a
ultraviolet cutoff at the Planck scale $\Lambda_p\simeq 10^{19}$GeV,
the
vacuum-energy density (\ref{0}) is of the order of $\Lambda_p^4\simeq
10^{76}$GeV. This
is $10^{123}$ times larger than the present observational data,
$10^{-47}$GeV, and it is unclear
how they incorporate  with each other unless an
extremal fine-tuning is made. This is 
the cosmological constant problem, as discussed in details by Weinberg
in \cite{wei68}, 
which challenges the fundamental 
theories, both the local quantum field theory and General Relativity.

The approaches to this problem can be briefly classified into
three categories: (i) fundamental physics, (ii) ``quintessence'' and
(iii) the anthropic 
principle. In the first category, all negative energy states are fully
filled and 
the mean-value of the vacuum-energy density is positive, the small
$\Lambda$-term is related to: 
the gravitational potential between virtual particles;
various scales of fundamental physics, such as electroweak processes,
inflationary particle
creations, et al.\cite{starob}. The elegant supersymmetry forces in
eq.(\ref{0}) 
bosonic and fermionic contributions to precisely cancel each other and
its breaking 
scale gives rise to a $\Lambda$-term. In the second category,
``quintessence''
\cite{gv,qts} postulates a new self-interacting scalar field $\Phi$
with a potential  $V(\Phi)$, 
incorporating 
within inflationary model and quantum cosmology\cite{cds}. In the
third category, the 
anthropic principle describes the probability of the $\Lambda$-term
conditioned by the necessity for the suitable evolution of intelligent

life\cite{weinberg}.

In this letter we present an alternative view on 
the problem of the cosmological constant and vacuum energy. 
Namely, the cosmological constant is arised 
only from the contributions from {\it the relevant 
vacuum fluctuations} whose wavelengths are conditioned by the {\it
size, 
homogeneity, geometry} and {\it topology}
of the Universe. 

\section{Relevant Vacuum Fluctuations}

The vacuum-energy density (\ref{0}), by the order of magnitude, 
does not strongly depend on the details of spectra $\epsilon_B({\bf
k})$ and 
$\epsilon_F({\bf k})$, i.e., massive or massless, free
or interacting one. Instead, it strongly depends on the number 
of high energy modes, 
since the vacuum-energy density (\ref{0}) is mainly 
contributed from the high energy modes of vacuum fluctuations of
various quantum fields at short distances.

In the general relativity, the gravity must be generated by all kinds 
of energy-mass, including the vacuum energy created by vacuum
fluctuations of various quantum
fields. However, it seems very 
mysterious why such a hugh vacuum-energy density $O(10^{76})$GeV
(\ref{0}), 
contributed from each spice of quantum fields,
is absent in the right-handed side of the Einstein equation, thus 
has no any effect on the classical gravity. 
It is conceivable that this fact could be due to two possibilities 
at the distance of the Planck length $L_{pl}$: (i) hidden symmetries
forcing 
the vacuum-energy density (\ref{0}) to be exactly zero; (ii) a certain
type 
of dynamical cancelations between 
quantum gravity's contributions and quantum field's contributions in 
the vacuum-energy density (\ref{0})\cite{xue00}.\footnote{In
\cite{xue00}
it is attempted to understand the possibility of the vacuum energy
cancellation due to the energy contribution from unstable modes of
quantum gravity.} 
We are not ambitious to cope with this problem here.  Instead, we
assume that the mean-value
of the vacuum-energy density (\ref{0}) is zero and does not contribute
to the right-handed side of the Einstein equation. 
  
On the other hand, it is equally unclear how 
the left-handed side of the Einstein equation describing the geometry 
and the large-scale Universe at present epoch can contain a
non-vanishing 
cosmological term $\Lambda g_{\mu\nu}$. This is the problem we would
like 
to address in this paper. We first define the notion of 
{\it the vacuum fluctuation}:
it is a causally-correlated fluctuation of the vacuum, upon the zero
mean-value
of the vacuum-energy density (\ref{0}). This has to be distinguished
from the 
fluctuations of various quantum fields in the vacuum, which contribute
to 
eq.(\ref{0}). 
The non-vanishing cosmological term $\Lambda g_{\mu\nu}$ is originated

from the relevant modes of the vacuum fluctuation. 

Consider a complex scalar field $\phi$ 
mimicing the causally-correlated vacuum 
fluctuations upon the zero
mean-value of the vacuum-energy density (\ref{0}).  
The simplest coordinate-invariant action $\tilde S$ for the quantum
scalar 
field $\tilde S$ is given by ($\hbar=c=1$)
\begin{equation}
\tilde S = {1\over2}\int
d^4x\sqrt{-g}\Big[g^{\mu\nu}\phi_{,\mu}\phi^*_{,\nu}+(m^2+\xi {\cal
R})\phi\phi^*\Big],
\label{action}
\end{equation}
where $m$ is an effective mass of the scalar field and $\xi$ is the  
coupling constant to the Riemann scalar ${\cal R}$. In terms of the
Riemann tensor
${\cal R}_{\mu\nu}$ and the energy-momentum tensor $T_{\mu\nu}$ 
of the classical matter, the Einstein equation is written as 
\begin{equation}
{\cal R}_{\mu\nu}-{1\over2}{\cal R}g_{\mu\nu}+ 
8\pi G\langle \tilde T_{\mu\nu}\rangle_r
=-8\pi GT_{\mu\nu}.
\label{e5'}
\end{equation}
The cosmological term is described by an averaged energy-momentum
tensor 
$\langle \tilde T_{\mu\nu}\rangle_r$ of the quantum scalar field
$\phi$:
\begin{equation}
\langle \tilde T_{\mu\nu}(x)\rangle_r = 
-{2\over\sqrt{-g}}{\delta \ln Z_r\over \delta
g^{\mu\nu}(x)},\hskip0.3cm 
Z_r=\langle 0|0\rangle_r=\int [{\cal D}\phi{\cal
D}\phi^*]_r\exp(-\tilde S),
\label{e1}
\end{equation}
which is averaged over the relevant modes of the vacuum fluctuation 
with the amplitude (the partition function $Z_r$) of transition
between
relevant modes of the vacuum fluctuation in the background 
of Einstein equation's solution $g_{\mu\nu}(x)$ and the global
topology of the Universe.

Given the action $\tilde S$ (\ref{action}) of the quantum scalar field
$\phi$ 
in the curved space-time of the Universe with its topology,  we can in
principle 
determine a unique complete and orthogonal basis of wave-functions 
$u^r_{\bf k}(x)$ of relevant modes $\bf k$ of the scalar field:    
\begin{equation}
\phi(x)=\sum_{\bf k}\Big(a_{\bf k}u^r_{\bf k}(x)+
a^\dagger_{\bf k}u^{r*}_{\bf k}(x)\Big),
\label{de}
\end{equation}
where $a_{\bf k}$ is the amplitude of the ${\bf k}$-th relevant mode.
On this relevant basis, the partition function $Z_r$ can be computed
as  
\begin{eqnarray}
Z_r&=&[\det\left(M^r\right)]^{-1},\hskip0.3cm
M^r_{\bf k,k'}=\int d^4x\sqrt{-g}u^r_{\bf k}(x)(\Delta_x + m^2 
+ \xi R)u^{r*}_{\bf k'}(x)\nonumber\\
\Delta_x &=&
{1\over\sqrt{-g}}\partial_\mu\big[\sqrt{-g}g^{\mu\nu}\partial_\nu\big]
.
\label{e1m}
\end{eqnarray}
Diagonizing the hermitian matrix $M^r$, we obtain  
\begin{equation}
\ln Z_r=-\int d^4x\sqrt{-g}\int{d^4k\over (2\pi)^4}\ln(\lambda^r_{\bf
k}),
\label{zr}
\end{equation}
where $\lambda^r_{\bf k}$ denotes the $\bf k$-th eigen-value of the 
matrix $M^r$. Thus, the averaged energy-momentum tensor 
$\langle \tilde T_{\mu\nu}\rangle_r$ (\ref{e1}) is given by,
\begin{equation}
\langle \tilde T_{\mu\nu}\rangle_r = g_{\mu\nu}(x)\int{d^4k\over
(2\pi)^4}
\ln (\lambda^r_{\bf k}),
\label{e1'}
\end{equation}
where we approximately neglect the functional variation $\delta
g_{\mu\nu}(x)$ 
of eigen-values $\lambda^r_{\bf k}$ in the logarithmical function.
We identify the cosmological constant, 
\begin{equation}
\Lambda=8\pi G \int{d^4k\over (2\pi)^4}\ln\lambda^r_{\bf k}.
\label{lambda}
\end{equation}
This clearly indicates that the cosmological constant is determined by
the eigen-values of relevant modes of the scalar field in the
background 
of the Einstein equation. 
 
In eq.(\ref{action}), the mass of the scalar 
field  $m$ is scaled by the masses $m_f$ of virtual fermions and 
anti-fermions that are annihilated and created in vacuum fluctuations
of quantum fields. 
Obviously, $m\ll \Lambda_p$, otherwise there would not be any vacuum 
fluctuations. Analogous to eq.(\ref{0}), major vacuum-fluctuation
modes 
contributing to the averaged energy-momentum tensor (\ref{e1'}) 
is stemming from the high-energy range $(m,\Lambda_p)$, and we can 
approximately neglect the mass term $m\phi^2$ in computations.
Analogously, 
we approximately neglect the local interacting term $\xi\phi^2 R$,
since the scale 
of the Riemann scalar ${\cal R}$ describing the large structure of the
Universe
is even very much smaller than $m$. One should not expect any
significant 
coupling between very rapid variations of high-energy modes at short
distances 
and the Riemann scalar ${\cal R}$ at long distances.

For action (\ref{action})
with $m=0$ and $\xi=0$, the massless scalar field $\phi$ obeys the
equation of motion,
\begin{equation}
M^r\phi=\Delta_xu^r_{\bf k}(x)=0,\hskip0.3cm 
u^r_{\bf k}(x)={\cal Y}^r_{\bf k}(r,\theta,\phi)\chi^r_{\bf k}(\eta).
\label{mode}
\end{equation}
The function ${\cal Y}^r_{\bf k}(r,\theta,\phi)$
fulfills the equation:
\begin{equation}
\Delta^3{\cal Y}^r_{\bf k}(r,\theta,\phi)=
-|{\bf k}|^2{\cal Y}^r_{\bf k}(r,\theta,\phi),\hskip0.3cm \Delta^3=
{1\over\sqrt{-h}}\partial_i\big[\sqrt{-h}h^{ij}\partial_j\big].
\label{y}
\end{equation}
The function $\chi^r_{\bf k}(\eta)$ obeys the equation
\begin{equation}
{\partial^2\chi^r_{\bf k}(\eta)\over\partial\eta^2}+
|{\bf k}|^2\chi^r_{\bf k}(\eta)=0,\hskip 0.3cm \chi^r_{\bf k}(\eta)
\sim e^{i\epsilon(|{\bf k}|)\eta}
\label{chi}
\end{equation}
where $\epsilon(|{\bf k}|)=|{\bf k}|$. The solution
$\chi^r_{\bf k}(\eta)
 \sim e^{i\epsilon(|{\bf k}|)\eta}$ shows the positive spectrum
$\epsilon(|{\bf k}|)$ of
the massless quantum scalar field with respect to the Killing vector 
$\partial_\eta$ for $\eta,t\rightarrow\infty$. 

\section{The Current Value of the Cosmological Constant}

In order to find the relevant modes in FRW Universe contributing to
the cosmological constant $\Lambda$ (\ref{lambda}), we have to
solve the eigenvalue equation (\ref{y}) for a
given size, homogeneity, geometry and topology of the Universe. We
will
consider
for simplicity the flat FRW Universe $K=0$, with the radius $a(t)$ and

topology $T \times R^3 $ 
where $T$ is the time and $R^3$ is the compactified spatial manifold 
described by the coordinates ($r,\theta,\phi$). The general 
solution of Eq.(\ref{y}) can be written as
\begin{equation}
{\cal Y}^r_{\bf k}(r,\theta,\phi)\sim j_l(k_rr)Y_{lm}(\theta, \phi),
\label{s3}
\end{equation}
where $Y_{lm}(\theta,\phi)$ is the spherical harmonic function and 
$j_l(k_rr)$ is the spherical Bessel
function with the radial momentum $k_r$ and the angular quantum number

$l=0,1,2...$. The eigen-value $\lambda^r_{\bf k}$ in eq.(\ref{zr}) is
then given by,
\begin{equation}
(\lambda^r_{\bf k})^2=k_t^2+k_r^2+{l(l+1)\over r^2},
\label{eigen}
\end{equation}
where $k_t$ is the temporal component of eigen-value $\lambda^r_{\bf
k}$. 
Integrating over $k_t$ in  eq.(\ref{lambda}) leads to 
\begin{equation}
\Lambda = 8\pi G \int{d^3k\over (2\pi)^3}\epsilon(|{\bf k}|), 
\hskip0.3cm \epsilon(|{\bf k}|)=\sqrt{k_r^2+{l(l+1)\over r^2}},
\label{lambda1}
\end{equation}
up to an irrelevant integral constant independent of $|{\bf k}|$. 
 
It is crucial that in our problem, due to the cosmological 
principle, the angular quantum number $l$ cannot be any other values 
except $l=0$ in the general solution of eq.(\ref{s3}). 
In the other words, due to the homogeneity and
isotropy of the Universe, there is
not a chosen point $r=0$ of the space-time where $\langle  \tilde
T(x)\rangle_r$ and $\phi(x)$ vanish, which just requires $l=0$.
Angular 
quantum numbers of relevant modes of the vacuum fluctuation $\phi$
must 
be zero. In addition, the quantum scalar field $\phi$ is
confined within the manifold of the topology $R^3$ and we have a
simple 
boundary value problem with the Dirichlet condition,
\begin{equation}
j_\circ(\alpha_\circ^n)=
0,\hskip 0.3cm k^n_r={\alpha_\circ^n\over a},
\label{s5}
\end{equation}
where $\alpha_\circ^n$ is the n-th zero-point of the spherical Bessel
function $j_\circ(x)$. The $k^n_r$ denotes the radial momentum of
relevant
modes contributing the cosmological constant $\Lambda$.
In our Universe $a(t)\gg 1$ and the asymptotic behavior of
$j_\circ(x)$ is
\begin{equation}
j_\circ(k_ra(t))\simeq {2\over a(t)}\sin(k_ra),
\label{s6}
\end{equation}
and we find $\alpha_\circ^n=k^n_ra(t)=n\pi,n=0,\pm 1,\pm
2,\cdot\cdot\cdot$, where $a(t)$ is the scale factor of the Universe.
The positive spectrum $\epsilon(|{\bf k}|)$ of relevant modes of the
vacuum 
fluctuation is given by,
\begin{equation}
\epsilon(k_r)=k_r={\pi n\over a},\hskip 0.3cm n=0,1,
2,\cdot\cdot\cdot .
\label{s7}
\end{equation}
As a result, we obtain the cosmological constant $\Lambda$,
\begin{equation}
\Lambda = 8\pi G \sum_l{(2l+1)\over a^2(t)}\int{dk_r\over (2\pi)}
\sqrt{k_r^2+{l(l+1)\over a^2}}=8\pi G {\pi\over a^4}
\sum _nn.
\label{lambda2}
\end{equation}

The $\Lambda$-density $\rho_\Lambda$ is then given by,
\begin{equation}
\rho_\Lambda\equiv {\Lambda\over 8\pi G}= {1\over2}{\hbar c\pi\over
2a^4}
N_{\max}(N_{\max}+1),
\label{011}
\end{equation}
where $N_{\max}$ is the maximum number of relevant
modes in the radial direction.
Let us estimate the present value of the $\rho_\Lambda$.
The maximum number of relevant modes in the radial direction is
approximately
given by
\begin{equation}
N_{\max}\simeq {a\over L_{pl}}\simeq 10^{61},
\label{nmax}
\end{equation}
where the present size of the Universe $a\simeq 1\cdot 10^{28}\, cm$
and
the Planck length is $L_{pl}\simeq 1.6 \, 10^{-33} \, cm$.
This numerically yields the present value of the $\Omega_\Lambda$
\begin{equation}
\rho_\Lambda = \frac{\hbar c \pi}{2a^2}L_{Pl}^{-2}\simeq 5.5\,
10^{-28}\, g \, {\rm cm}^{-3},
\label{omega}
\end{equation}
consistently with recent observations. 
The corresponding vacuum pressure is negative and thus accelerates
expansion of the Universe.
 
 The proportionality coefficient in  
(\ref{011}) 
of geometry and topology of the Universe, although computations must
be
more complicated. An alternative approach avoiding the difficulties of
the eigenvalue 
problem in hyperbolic manifolds is the theory of dynamical systems
used for the study of the properties of CMB \cite{gk1}. 
 
The probability of creation of the
Universe within the framework of quantum cosmology depends not only on
the
matter fields but also on the cosmological constant and the topology.
Particularly, the $S^3$ topology has the highest probability
determined
via the wave function of the Universe, among the
considered $R\times S^3$, (K=1), $R\times H^3/\Gamma$ (K=-1) $R^4$
(K=0)
topologies \cite{gk} and $T\times S^3$, (K=0)\cite{xue88}. In the case
of 
inflationary Universe, the probabilities for the positively and
negatively curved and
flat geometries become comparable.

\section{Discussion}

Thus, in our approach the cosmological term in the left-handed side of
the 
Einstein equation describing the present Universe is attributed to the
vacuum 
fluctuation $\phi$ whose wavelengths are conditioned by the size,  
homogeneity, geometry and topology of the Universe. Due to the small
anisotropy of 
the Universe, the angular quantum numbers $l$ of the relevant modes of
the vacuum 
fluctuations (\ref{lambda2}) could be finite numbers.  As a
consequence, the
cosmological constant would be angular dependent, however not more
than
the order of the anisotropy of the Cosmic Background radiation,
$10^{-5}$. 
 
The density (\ref{011}) is in fact an 
{\it absolute} value of the energy density of the vacuum fluctuations
(\ref{e1'}).
If it was the energetic difference of the energy density (\ref{e1'}) 
with and without presence of the classical gravity, 
the resulted cosmological term would be 
$\sim\hbar c/a^4$, which can be obtained by analogous computations
leading 
to the Casimir effect \cite{xue01}.  

The remarkable numerical coincidence of the obtained density with that
of indicated by the supernovae observations, is prompting the idea
of the variation of the Planck length by the expansion of the
Universe,
or
\begin{equation}
\frac{G}{c^4} = {\rm const}\cdot a^{-2},
\label{g}
\end{equation}
which is actually in the line of Dirac's old idea of the time
variation of the gravitational constant.

If this condition is fulfilled, the cosmological constant will remain
constant, and its
equality with the present matter density would be a chance
coincidence.
That is, the cosmological constant would be smaller than the matter
density
in the past and higher in the future. The absence of the contribution
of the modes with non-zero angular quantum numbers with other
formulations of the cosmological principle seems also a remarkable
fact and needs further studies. 
The mechanism we discussed, will thus indicate
the fine tuning between the macro and micro structure of the Universe,
the nature of which must follow from future fundamental theories.

\end{document}